\documentclass[%
reprint,
superscriptaddress,
 amsmath,amssymb,
 aps,
prb,
]{revtex4-1}

\usepackage{graphicx}
\usepackage{dcolumn}
\usepackage{bm}
\usepackage{hyperref}
\usepackage{sidecap}
\usepackage{float}

\begin{document}
\title{Spatially resolved penetration depth measurements and vortex manipulation in the ferromagnetic superconductor ErNi$_2$B$_2$C}
\author{Dirk Wulferding}
\altaffiliation{Contributed equally to this work}
\affiliation{Center for Artificial Low Dimensional Electronic Systems, Institute for Basic Science, 77 Cheongam-Ro, Nam-Gu, Pohang 790-784, Korea}
\affiliation{Department of Physics, Pohang University of Science and Technology, Pohang 790-784, Korea}

\author{Ilkyu Yang}
\altaffiliation{Contributed equally to this work}
\affiliation{Center for Artificial Low Dimensional Electronic Systems, Institute for Basic Science, 77 Cheongam-Ro, Nam-Gu, Pohang 790-784, Korea}
\affiliation{Department of Physics, Pohang University of Science and Technology, Pohang 790-784, Korea}

\author{Jinho Yang}
\affiliation{Center for Artificial Low Dimensional Electronic Systems, Institute for Basic Science, 77 Cheongam-Ro, Nam-Gu, Pohang 790-784, Korea}
\affiliation{Department of Physics, Pohang University of Science and Technology, Pohang 790-784, Korea}

\author{Minkyung Lee}
\affiliation{Center for Artificial Low Dimensional Electronic Systems, Institute for Basic Science, 77 Cheongam-Ro, Nam-Gu, Pohang 790-784, Korea}
\affiliation{Department of Chemistry, Pohang University of Science and Technology, Pohang 790-784, Korea}

\author{Hee Cheul Choi}
\affiliation{Center for Artificial Low Dimensional Electronic Systems, Institute for Basic Science, 77 Cheongam-Ro, Nam-Gu, Pohang 790-784, Korea}
\affiliation{Department of Chemistry, Pohang University of Science and Technology, Pohang 790-784, Korea}

\author{Sergey L. Bud'ko}
\affiliation{Ames Laboratory, U.S. Department of Energy, Ames, Iowa 50011, USA}
\affiliation{Department of Physics and Astronomy, Iowa State University, Ames, Iowa 50011, USA}

\author{Paul C. Canfield}
\affiliation{Ames Laboratory, U.S. Department of Energy, Ames, Iowa 50011, USA}
\affiliation{Department of Physics and Astronomy, Iowa State University, Ames, Iowa 50011, USA}

\author{Han Woong Yeom}
\affiliation{Center for Artificial Low Dimensional Electronic Systems, Institute for Basic Science, 77 Cheongam-Ro, Nam-Gu, Pohang 790-784, Korea}
\affiliation{Department of Physics, Pohang University of Science and Technology, Pohang 790-784, Korea}

\author{Jeehoon Kim}
\email[]{Corresponding author: jeehoon@postech.ac.kr}
\affiliation{Center for Artificial Low Dimensional Electronic Systems, Institute for Basic Science, 77 Cheongam-Ro, Nam-Gu, Pohang 790-784, Korea}
\affiliation{Department of Physics, Pohang University of Science and Technology, Pohang 790-784, Korea}

\date{\today}

\begin{abstract}
We present a local probe study of the magnetic superconductor, ErNi$_2$B$_2$C, using magnetic force microscopy at sub-Kelvin temperatures. ErNi$_2$B$_2$C is an ideal system to explore the effects of concomitant superconductivity and ferromagnetism. At 500 mK, far below the transition to a weakly ferromagnetic state, we directly observe a structured magnetic background on the micrometer scale. We determine spatially resolved absolute values of the magnetic penetration depth $\lambda$ and study its temperature dependence as the system undergoes magnetic phase transitions from paramagnetic to antiferromagnetic, and to weak ferromagnetic, all within the superconducting regime. In addition, we estimate the absolute pinning force of Abrikosov vortices, which shows a position- and temperature dependence as well, and discuss the possibility of the purported spontaneous vortex formation.

\begin{description}

\item[PACS numbers]{74.25.Wx, 74.25.Ha, 68.37.Rt}

\end{description}
\end{abstract}

\maketitle

\section{Introduction}

The interplay between magnetism and superconductivity has caught the attention of many condensed matter scientists. Once thought to be mutually exclusive, several materials have since emerged in which a coexistence of magnetic order and superconductivity can be found.~\cite{canfield-98, lynn-97, muller-01} This is contrasted, e.g., by high temperature cuprate superconductors; although a low temperature rare-earth-related magnetic ordering may occur within the superconducting phase,~\cite{simizu-87} there is a clear separation between the Cu-based strong antiferromagnetic and the superconducting phases.~\cite{scalapino-12} Nevertheless, spin fluctuations in the Cu-O planes that arise from the melting of the antiferromagnetically ordered phase upon doping seem to support and mediate superconductivity.~\cite{scalapino-12, wulferding-14}

The possibility of studying materials that exhibit magnetic order and superconductivity simultaneously could shed a new light on the importance of magnetic correlations for the pairing mechanism of Cooper pairs in unconventional superconductors. In particular, our insight into high temperature superconductivity could greatly benefit from a detailed study of the interplay among intrinsic magnetism, penetration depth $\lambda$, and coherence length $\xi$.

From an applied point of view, the pinning of Abrikosov vortices in type-II superconductors is a central issue, as increased pinning forces $F_P$ can greatly enhance both critical currents $J_c$ and critical magnetic fields $H_{c2}$.~\cite{blatter-94} The presence of an intrinsic magnetic field (with $B_{int} < H_{c2}$) can have a strong effect on $F_P$ and may thus present an alternative to artificially introduced pinning centers.

The rare earth borocarbides $R$Ni$_2$B$_2$C (with $R$ = Dy, Ho, Er, and Tm) exhibit magnetism within the superconducting phase.~\cite{canfield-98} Conspicuously, their superconducting and magnetic transition temperatures are comparable, suggesting similar energy scales for magnetism and superconductivity.~\cite{muller-01} The observation of a pronounced isotope effect in these intermetallic compounds points towards conventional phonon mediated superconductivity.~\cite{lawrie-94, cheon-99} Of these borocarbides, ErNi$_2$B$_2$C has a relatively high superconducting transition temperature of $T_c = 10.5$ K and develops antiferromagnetic order around $T_N = 6$ K.~\cite{cho-95} Remarkably, below $T_{WFM} = 2.3$ K a weak ferromagnetic phase can be detected with a net magnetic moment of 0.39$\mu_B$/Er and a resulting intrinsic magnetic field of about $B_{int} = 500 - 700$ G.~\cite{canfield-96, bluhm-06} Previous tunnel diode oscillator (TDO) experiments focusing on the magnetic penetration depth $\lambda$ uncovered a pronounced temperature dependence with clear features at $T_N$ and $T_{WFM}$, therefore indicating a strong impact of the magnetic properties on $\lambda$, as well as fingerprints of the formation of a spontaneous vortex lattice (i.e. induced by intrinsic magnetism) below $T_{WFM}$.~\cite{chia-06} At the same time, Bitter decoration experiments~\cite{vinnikov-05} and scanning Hall probe measurements~\cite{bluhm-06} revealed a microscopic variation of the intrinsic magnetic field emerging below $T_N$ in the form of well-ordered, micrometer sized magnetic stripes, along which vortices tend to order. A spontaneous vortex phase in the weak ferromagnetic regime was not observed, though. However, the Bitter decoration experiments were conducted down to 1.9 K, which is only slightly below $T_{WFM}$, and the spatial resolution of scanning Hall probes is rather limited. In fact, until now no clear consensus on the issue of spontaneous vortex lattice formation in ErNi$_2$B$_2$C has been reached, partly due to its low ferromagnetic transition temperature. This open issue together with the microscopic variation of the magnetic field calls for a local probe investigation of ErNi$_2$B$_2$C at sub-Kelvin temperatures.

Here we present a magnetic force microscopy (MFM) study on ErNi$_2$B$_2$C at He-3 temperatures and with a high spatial resolution. We observe magnetic stripe-like features at $T=500$ mK, that arise from twin domain boundaries. Their pronounced temperature behavior supports the spontaneous vortex formation scenario. Using the magnetic moment of the tip we estimate the pinning force of a single vortex. In addition, a spatially resolved determination of the penetration depth $\lambda$ is presented, which reveals a close relation between $\lambda$ and the microscopically varying intrinsic magnetism.

\section{Experimental details}

A single crystal of ErNi$_2$B$_2$C was grown via Ni$_2$B flux growth technique.~\cite{cho-95, canfield-crystal} The sample's magnetic properties have been measured to determine $T_c$ using an MPMS superconducting quantum interference device (Quantum Design, Inc.). The crystal orientation was determined by polarized Raman scattering experiments, using the orientation dependence of the $B_{1g}$ phonon mode intensity.~\cite{drachuck-12} Prior to MFM measurements, the sample was mechanically polished to obtain a fresh surface within the $ab$ plane with a roughness as small as 5 nm. MFM measurements were performed with a unique home-built He-3 MFM probe with a vector magnetic field capability (2T -- 2T -- 9T along the $x$-$y$-$z$ direction) and a temperature range of 320 mK -- 300 K.~\cite{jinho-mfm} All experiments were carried out with no magnetic fields applied and with a commercially available tip.~\cite{mfm-tip}

\section{Experimental Results $\&$ Discussion}

\begin{figure*}
\label{figure1}
\centering
\includegraphics[width=17.5cm]{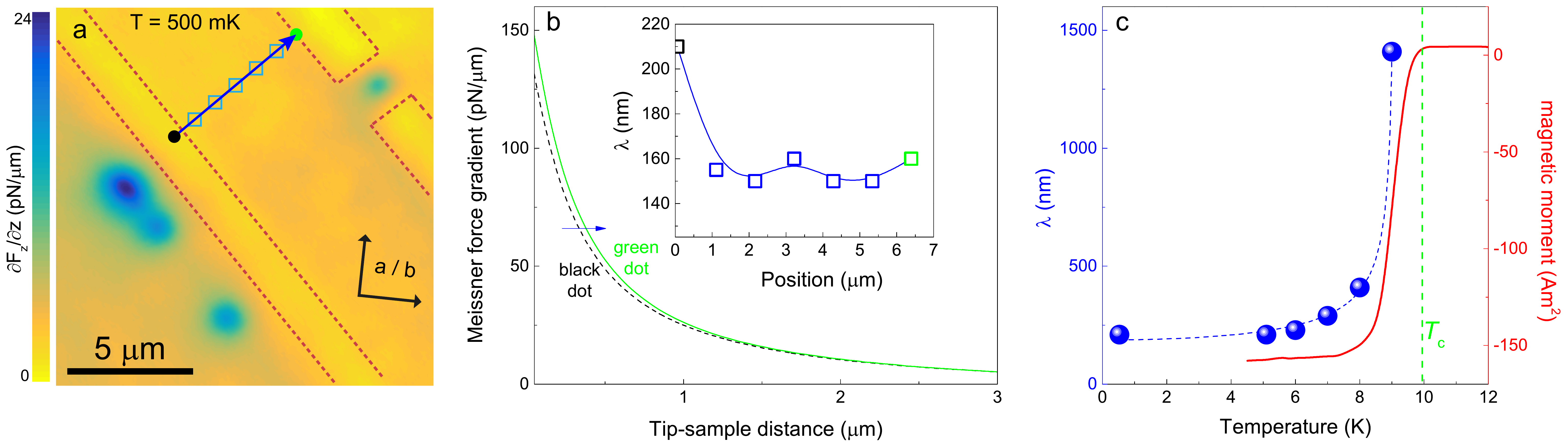}
\caption{(Color online) (a) A 15 $\mu$m $\times$ 15 $\mu$m MFM image obtained at $T = 500$ mK and 300 nm tip-sample distance. The blue arrow marks a line of 6.5 $\mu$m length along which spatially resolved $\lambda$ measurements were performed. The dashed lines outline the magnetic stripes. (b) Meissner force curves obtained at the beginning [i.e. black dot in (a)] and at the end of the arrow [i.e. green dot in (a)]. The inset shows local $\lambda$ values along the blue arrow in (a). (c) Temperature dependence of $\lambda$ obtained at a single position (blue spheres) and the magnetic moment measured at $H = 30$ Oe (solid red line). The dashed blue line is a fit according to the BCS theory.}
\end{figure*}

We first discuss the origin of magnetic stripes and the local variation of the magnetic penetration depth, before moving on to the pinning force. Synchrotron x-ray scattering experiments on ErNi$_2$B$_2$C revealed a structural tetragonal-to-orthorhombic distortion upon entering the antiferromagnetic phase at $T_N = 6$ K. This leads to the formation of twin domain boundaries below $T_N$ along [110] and [1$\bar{1}$0],~\cite{detlefs-97} which are already inhabited by a ferromagnetic component aligned along the $c$ axis.~\cite{saha-00} In small applied magnetic fields (of the order of 10 G) vortices were found along the boundaries, thus forming stripes with roughly 5 $\mu$m spacing.~\cite{veschunov-07} The transition at $T_{WFM}$ is driven by magneto-elastic coupling, which enhances Ruderman-Kittel-Kasuya-Yosida (RKKY) interactions among the Er $4f$ electrons, and thereby a long-range ordered weak ferromagnetism is induced in the bulk, as evidenced by inelastic neutron scattering experiments.~\cite{kawano-02} In our MFM data [see Fig. 1(a)] a structured magnetic background can be clearly seen in the form of bright stripes with a separation of $\approx$ 7.5 $\mu$m at $T=500$ mK, i.e., deep within both the superconducting and the weak ferromagnetic phase. The stripes run from the lower right corner to the upper left corner [highlighted by dashed lines in Fig. 1(a)] and at 45$^{\circ}$ from the crystallographic $a$ and $b$ axes. We therefore ascribe the stripes to regions of increased ferromagnetism. Their orientation along the [110] or [1$\bar{1}$0] direction suggests a coincidence with twin domain boundaries.~\cite{detlefs-97, veschunov-07} Indeed, magnetic stripes of the same orientation and comparable spacing have been observed in Bitter decoration experiments for $T<T_N$ and attributed to twin boundaries.~\cite{saha-00, vinnikov-05} However, the magnetic stripes in our experiment evidence a considerably different temperature behavior, as discussed further below.

In order to probe the influence of the varying internal magnetic field on the superconductivity in the sample, we performed local measurements of $\lambda$ across two stripes along the blue arrow in Fig. 1(a). The determination of $\lambda$ can be done by slowly bringing the tip to the surface at a fixed sample position and measuring the Meissner force gradient acting upon the magnetic tip. By comparing the Meissner force curves to those of a well-known standard sample (a Nb film), one can obtain the absolute values of $\lambda$.~\cite{kim-12} In Fig. 1(b) we plot two Meissner force curves measured at the beginning (dashed black) and at the end (solid green) of the arrow. A clear shift is observed, indicating a spatial variation in $\lambda$. In the inset of Fig. 1(b) several absolute $\lambda$ values at different positions (open squares) are plotted. These values, measured under the same conditions, vary by more than 20$\%$ within a few micrometers, while following closely the changes in the intrinsic ferromagnetism due to the stripe formation.~\cite{asymmetry-comment} This is to be expected, as the magnetic penetration depth will be enhanced in the presence of an increased intrinsic magnetic field. In fact, we can reinterpret our MFM images in the superconducting state as maps of changes in $\lambda$.

The temperature dependence of $\lambda$ [see Fig. 1(c)] can be well-described by a BCS-type behavior, i.e., $\lambda(T) = \frac{\lambda(0)}{\sqrt{1-(T/T_c)^2}}$, where $\lambda(0)$ is the penetration depth at $T=0$ K. The fit yields $\lambda(0) \approx 190$ nm and $T_c \approx 9.1$ K. Our directly obtained penetration depth $\lambda(0)$ is larger than those previously reported from bulk probe experiments,~\cite{yaron-96, cho-95} while $T_c$ is slightly below the reported value of 10.5 K. Aside from the different experimental approaches, it is known that the $T_c$ of as-grown samples is lower than that of annealed samples.~\cite{miao-02} Indeed, bulk magnetization measurements of our as-grown sample yield a $T_c$ of 9.9 K [see red line in Fig. 1(c)]. Another important consideration is a possible deviation of ErNi$_2$B$_2$C from the BCS behavior. In fact, a pronounced non-BCS-type behavior is apparent below $T_c$ in the highly resolved temperature dependence of the TDO study,~\cite{chia-06} which is related to the different magnetic phases within the superconducting regime. Since our temperature dependence is more coarse, a deviation from the BCS behavior could be obscured.

In addition to the magnetic stripes, we clearly see four separate vortices in Fig. 1(a). They are aligned antiparallel to the magnetic stripes, as indicated by the MFM contrast. Furthermore, they remain above $T_{WFM}$ and even above $T_N$ (as discussed below). Therefore, they exist independently of any intrinsic magnetic property. We ascribe these features to vortices induced by a weak external stray field. As this stray field is present regardless of the temperature (i.e., already above $T_c$, in contrast to the weak intrinsic ferromagnetism), its flux lines will be trapped inside ErNi$_2$B$_2$C upon cooling through $T_c$. The observation of $n = 4$ vortices on a scan area of $A = 15 \times 15$ $\mu$m$^2$ could already be accounted for by $H_{stray} = \frac{n \cdot \Phi_0}{A} \approx 0.37$ Oe, considering that each vortex corresponds to a flux quantum of $\Phi_0$. Although this estimation is somewhat rough due to our limited scan area, we expect that the stray field does not exceed 1 Oe, and is therefore below the accuracy of our vector magnet with a resolution of 1 Oe.

\begin{SCfigure*}
\label{figure2}
\centering
\includegraphics[width=12.5cm]{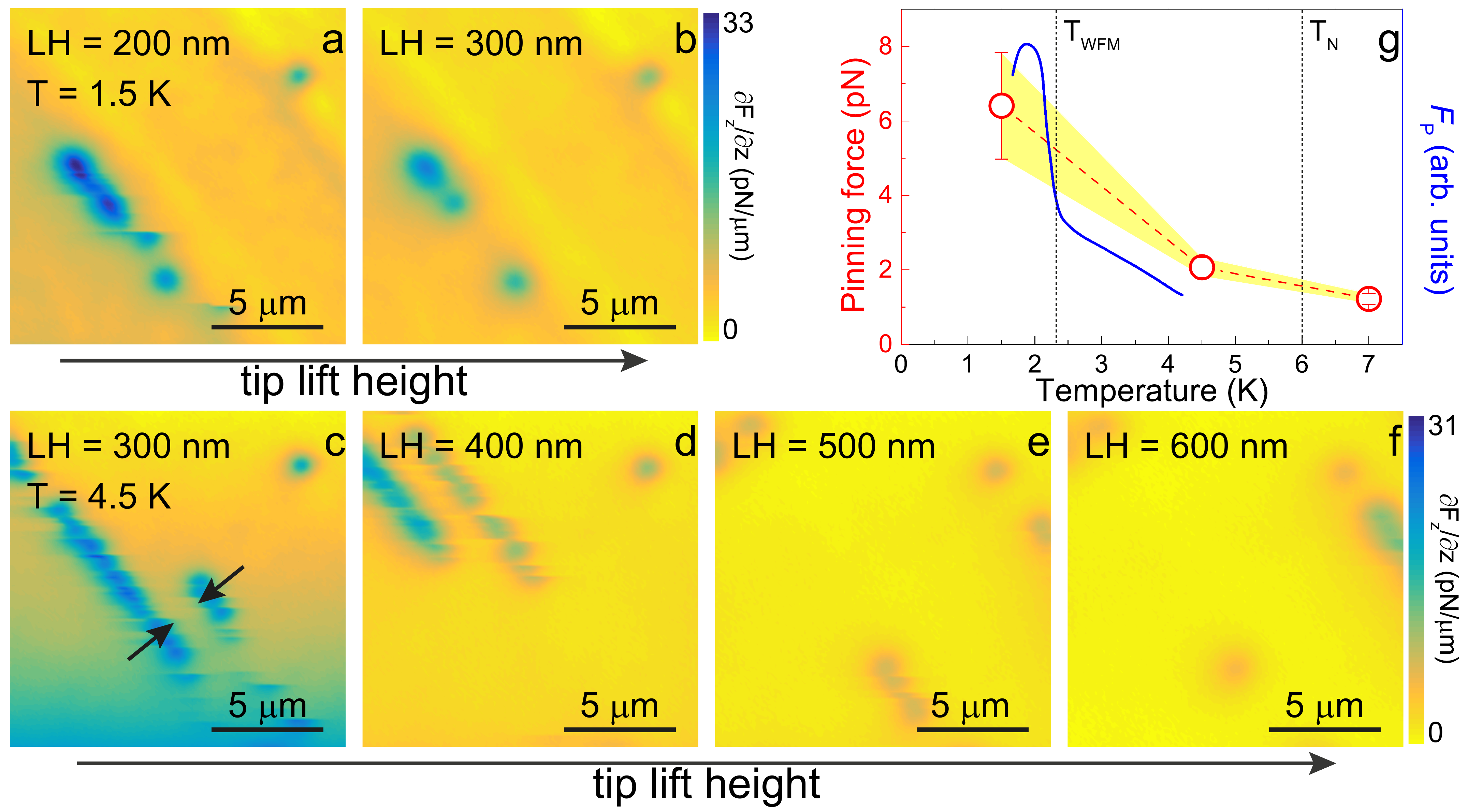}
\caption{(Color online) (a) and (b): MFM images taken at $T = 1.5$ K with 200 nm and 300 nm tip lift height ($LH$). (c) -- (f): MFM images taken at $T = 4.5$ K with 300 nm, 400 nm, 500 nm, and 600 nm tip lift height. (g) The pinning force of the vortices obtained from MFM measurements as a function of temperature (red circles) compared to the pinning force estimated from measurements of the critical current $I_c$ (solid blue line).~\cite{gammel-00}}
\end{SCfigure*}

In order to study the depinning of vortices in ErNi$_2$B$_2$C, we manipulate the separate, stray field induced vortices via the tip-sample interaction. In Figs. 2(a) and 2(b) two 15 $\mu$m $\times$ 15 $\mu$m MFM images are shown at $T = 1.5$ K and lift heights ($LH$) of 200 nm and 300 nm. The scan area corresponds to the one in Fig. 1(a). While at $LH=200$ nm a manipulation of the lower vortices along the twin boundary is observed, an increase to $LH=300$ nm leads to an image of static vortices. For all images the fast scan direction is horizontal, while the slow scan direction is vertical, from the bottom to the top. Note that the direction of the manipulated vortex motion always follows the bright stripe. We recall that the magnetic flux direction within the twin boundary is anti-parallel to that of the stray field vortices, as evidenced by the opposing contrast that correspond to opposing magnetic forces acting on the tip. It is therefore energetically unfavorable for a separate vortex to move into the stripe. On the other hand, the upper right vortex caught between two domain boundaries can move neither up nor down and hence experiences a much stronger pinning. To estimate the pinning force of vortices as a function of temperature, we consider the magnetic force between tip and vortex, given by $F_{tv} = \frac{m \Phi_0}{2 \pi} \cdot \frac{1}{[z + \lambda(T)]^2}$.~\cite{auslaender-08} Here, $m = (3.7 \pm 0.2)$ nAm is the magnetic moment per unit length of the tip, $\Phi_0 = h/2e$ is the magnetic flux quantum of a single vortex, and $z$ is the distance between tip and sample surface around which the vortex manipulation sets in. $m$ has been estimated via monopole-monopole approximation.~\cite{auslaender-08} Considering that at $T=1.5$ K the onset of the vortex manipulation is at a tip-sample distance of 200 nm -- 300 nm, we can get a rough estimation of the pinning force $F_P$ for vortices in the range of 5 -- 8 pN.

When increasing the temperature from 1.5 to 4.5 K and thereby moving from the weak ferromagnetic phase into the antiferromagnetic phase, the magnetic stripes vanish while the twin boundaries remain [see Figs. 2(c) -- 2(f)]. Hence, the background in the MFM images appears to be homogeneous. In contrast, stray field induced vortices remain as expected. In this intermediate temperature regime the pinning force is strongly reduced compared to the weak ferromagnetic phase. This is in good accordance with previous magnetization and transport experiments, where a threefold decrease in the relative $F_P$ was reported when heating the sample up through $T_{WFM}$ [see solid blue line in Fig. 2(g)].~\cite{gammel-00} Still, the direction of vortex manipulation remains along the twin boundaries, which continue to carry a reduced ferromagnetic component along the $c$ axis and anti-parallel to the vortex flux. This agrees well with the previous Bitter decoration experiment, where field-induced vortex stripes occur along [110] and [1$\bar{1}$0].~\cite{vinnikov-05} Their work reports parallel vortex stripes with 1 -- 2 $\mu$m spacing, which is also confirmed by our observations of parallel vortex lines aligned on both sides of the twin boundaries [see arrows in Fig. 2(c)]. We manage to manipulate the vortex position at 4.5 K up to a tip-sample distance of 500 nm. Only at $LH \geq 600$ nm, a static vortex image is obtained, leading to a pinning force of 2.0 -- 2.5 pN.

\begin{figure*}
\label{figure3}
\centering
\includegraphics[width=16cm]{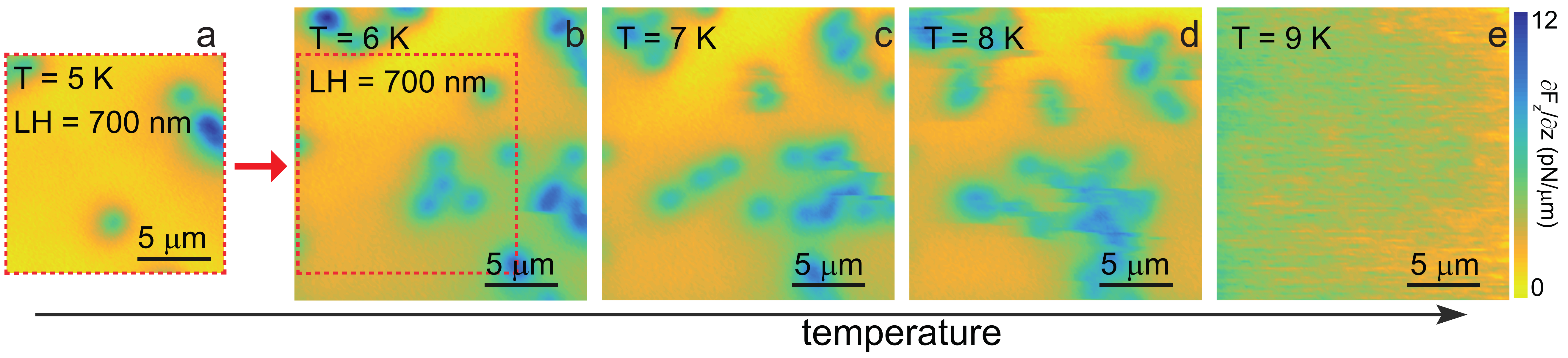}
\caption{(Color online) (a) A 15 $\mu$m $\times$ 15 $\mu$m MFM image obtained at $T=5$ K, (b) -- (e) 20 $\mu$m $\times$ 20 $\mu$m MFM images taken at $T$ = 6 K, 7 K, 8 K, and 9 K. The lift height for all images was constant at 700 nm.}
\end{figure*}

Moving from the magnetically ordered ($T < T_N$) into the paramagnetic superconducting regime ($T > T_N$), we study the vortex dynamics as a function of temperature at a constant $LH$ of 700 nm (see Fig. 3). It is important to note that the vortex configuration changes strongly around $T_N=6$ K [see Figs. 3(a) and 3(b)]. A detailed investigation of the critical current revealed a dynamic domain formation and annihilation in a narrow temperature regime around $T_N$, which leads to a considerable increase in antiferromagnetic antiphase domain boundaries on which vortices strongly get pinned.~\cite{weigand-13} These domain dynamics are evidenced by our observation of vortex rearrangements at $T = 6$ K. Above $T_N$, the structure relaxes from orthorhombic to tetragonal, leading to a disappearance of the twin boundaries, and hence to a random distribution of vortices. This is contrasted by the Bitter decoration study,~\cite{vinnikov-05} where a clear hexagonal vortex lattice forms for $T > T_N$ in an applied field of 25 Oe. Since our stray field is too weak to induce a considerable vortex density, local impurities may dominate the vortex arrangement instead. The increasing thermal energy of the vortices allows a manipulation with a $LH$ of 700 nm at 7 K and higher, as it leads roughly to a twofold reduction in the pinning force compared to $T = 4.5$ K. We estimate a pinning force at 7 K of around 1.0 -- 1.5 pN. The temperature dependence of the pinning force $F_P$ is plotted in Fig. 2(g). It mirrors the critical current behavior, which is closely related to $F_P$,~\cite{weigand-13} as both exhibit roughly a twofold decrease between 4.5 and 7 K. Moreover, our $F_P$ data is consistent with the observation of a finite $J_c$ at $T > T_N$ in Ref. [~\cite{weigand-13}], which was previously conjectured to vanish above the N\'{e}el temperature.~\cite{canfield-01} At 7 and 8 K in the absence of twin boundaries, there is no more preferred vortex manipulation direction. Instead, vortices are being dragged equally to the left and right while a general upwards trend in motion is observed, thus following the slow scan direction. Eventually, thermally enhanced vortex dynamics overcome the pinning potential and ErNi$_2$B$_2$C transits into a vortex liquid phase between 8 K and 9 K, i.e. close to $T_c$.

Our MFM images also allow us to comment on the issue of the coexistence mechanism in ErNi$_2$B$_2$C. It is believed that magnetism and superconductivity can coexist under two possible scenarios: ($i$) a spontaneous formation of vortices~\cite{greenside-81, kuper-80} or ($ii$) a variation of magnetic moments on a length scale smaller than the penetration depth $\lambda$.~\cite{greenside-81, blount-79} We clearly detect a modulation of the magnetic background over several micrometers. However, additional small-scale magnetic variations are not observed in our MFM images down to our resolution limit of 20 nm. Hence we do not find direct evidence for scenario ($ii$) from our data, although we cannot exclude possible magnetic variations on a smaller length scale than the instrumental limit. In addition, a phase separation between purely ferromagnetic and superconducting regions is not evident as the magnetic stripes still exhibit a Meissner force. On the other hand, scenario ($i$) is possible if the condition $H_{c1} < B_{int} < H_{c2}$ is fulfilled.~\cite{ng-97} For ErNi$_2$B$_2$C, the critical fields $H_{c1}$ and $H_{c2}$ are around 500 Oe and 12 kOe (below $T_{WFM}$), respectively.~\cite{canfield-96} As $B_{int} = 500 - 700$ G,~\cite{canfield-96, bluhm-06} we have a system at hand in which the spontaneous formation of vortices could very well be realized at $T < T_{WFM}$. We can estimate the number of vortices $n$ expected to appear in the presence of the intrinsic magnetic field. For a $15 \times 15$ $\mu$m$^2$ image as shown in Fig. 1(a) our estimation yields $n = \frac{B_{int} \cdot A}{\Phi_0} \approx 7600$ vortices. It is tempting to reinterpret the magnetic stripes below $T_{WFM}$ as regions of ultra-high vortex density, where single vortices cannot be resolved separately anymore. Considering that the twin boundaries carry a ferromagnetic component aligned along the $c$ axis,~\cite{saha-00} and that the flux lines of the vortices induced by the weak intrinsic ferromagnetism run along the $c$ axis as well, an accumulation of vortices in the vicinity of the twin boundaries could be energetically favorable. In this case the locally probed $\lambda$ values in the striped region [see black / green dot, Fig. 1(a)] do not reflect the true penetration depth, as each force gradient curve has a contribution of both the Meissner force from the superconducting region as well as the magnetic force from the flux quanta. In a previous Bitter decoration study vortex stripes pinned along the [100] direction have been observed at temperatures slightly below $T_{WFM}$ and in applied magnetic fields of 15 -- 20 Oe, which rearrange into a hexagonal vortex lattice above $T_{WFM}$.~\cite{veschunov-07} They were interpreted as a pinning of external field-induced vortices along ferromagnetic domains, that vanish above $T_{WFM}$ [see Figs. 2(b) and 2(c)]. A spontaneous flux lattice, on the other hand, was not observed. In our experiments, carried out in negligible external magnetic fields, the stripes vanish completely above $T_{WFM}$. This is in stark contrast to several previous Bitter decoration experiments~\cite{vinnikov-05, saha-00, veschunov-07} and further support for our scenario of the \textit{spontaneous} vortex formation induced by the weak intrinsic ferromagnetic field. Note that also a slight rearrangement of stray field induced vortices can be found upon crossing $T_{WFM}$, which could result from an interaction with departing spontaneous vortices. In order to unambiguously prove the existence of spontaneous vortex stripes in ErNi$_2$B$_2$C, however, further experiments with a higher resolution or within a lower vortex density region (e.g., by applying an opposing magnetic field) are required in a narrow temperature regime around the weak ferromagnetic transition.

\section{Summary}

In summary, we have presented an MFM study of the magnetic superconductor ErNi$_2$B$_2$C down to sub-Kelvin temperatures. We introduced a straightforward method for locally determining the vortex pinning force $F_P$, as well as demonstrated a spatially resolved measurement of the penetration depth $\lambda$. Our measurements yield $\lambda(0) \approx 190$ nm. Below $T_{WFM}$, $\lambda$ varies locally by more than 20$\%$ on a micrometer scale, while following closely changes in the intrinsic magnetic field. The pinning force of stray field induced vortices shows a strong temperature dependence that is related to the existence of various magnetic phases. We estimate the pinning forces to be around 1 -- 8 pN, depending on the temperature. At the same time we find that $F_P$ varies locally, and easy vortex manipulation directions exist in the magnetic phases, but are absent above $T_N$. Finally, a micrometer scale magnetic background detected at $T$=500 mK that coincides with the twin domain boundaries and vanishes above $T_{WFM}$ supports the scenario of spontaneous vortex stripe formation in ErNi$_2$B$_2$C.

\begin{acknowledgments}
We acknowledge important discussions with O. E. Ayala-Valenzuela. This work was supported by the Institute for Basic Science (IBS), Grant No. IBS-R014-D1. Work done by PCC and SLB was supported by the U.S. Department of Energy, Office of Basic Energy Science, Division of Materials Sciences and Engineering. Their research was performed at the Ames Laboratory. Ames Laboratory is operated for the U.S. Department of Energy by Iowa State University under Contract No. DE-AC02-07CH11358.
\end{acknowledgments}


\begin{thebibliography}{9}
\bibitem{asymmetry-comment}
Note, that a discrepancy between $\lambda$ values obtained at the start (black dot) and end point (green dot) might be related to a small hysteresis behavior of the piezo scanners.
\end{thebibliography}
\end{document}